\documentstyle{cupconf}

\ifoldfss
\else
  \ifnfssone
    \newmathalphabet{\mathit}
      \addtoversion{normal}{\mathit}{cmr}{m}{it}
      \addtoversion{bold}{\mathit}{cmr}{bx}{it}
    \newmathalphabet{\mathcal}
      \addtoversion{normal}{\mathcal}{cmsy}{m}{n}
    \else
    \ifnfsstwo
    \fi
  \fi
\fi

\def\upi{\pi} 
\def\hexnumber#1{\ifcase#1 0\or1\or2\or3\or4\or5\or6\or7\or8\or9\or
 A\or B\or C\or D\or E\or F\fi }

\makeatletter
\ifx\CUP@mtlplain@loaded\undefined
\else
\fi
\makeatother
 \makeatletter
 \ifx\CUP@mtlplain@loaded\undefined
   \font\tenbmi=cmmib10 at 10pt
   \font\sevenbmi=cmmib10 at 7pt
   \font\fivebmi=cmmib10 at 5pt

   \newfam\bmifam
   \textfont\bmifam=\tenbmi
   \scriptfont\bmifam=\sevenbmi
   \scriptscriptfont\bmifam=\fivebmi
   
 \fi
 \makeatother
\ifnfsstwo

\fi
\ifnfssone

\fi
\ifoldfss

\fi

\mathchardef\varLambda="0103

\makeatletter
\ifx\CUP@mtlplain@loaded\undefined
\else
\fi
\makeatother
\makeatletter
\ifx\CUP@mtlplain@loaded\undefined
  \font\tenbms=cmbsy10
  \font\sevenbms=cmbsy10 at 7pt
  \font\fivebms=cmbsy10 at 5pt
  \newfam\bmsfam
  \textfont\bmsfam=\tenbms
  \scriptfont\bmsfam=\sevenbms
  \scriptscriptfont\bmsfam=\fivebms

  \edef\bsy@{\hexnumber\bmsfam}
  \mathchardef\bnabla="0\bsy@72
\fi
\makeatother
\title[The Origin of Matter in the Universe]{
The Origin of Matter in the Universe: Reheating after Inflation}

\author[Lev Kofman]%
{L\ls E\ls V\ns  A.\ns K\ls O\ls F\ls M\ls A\ls N\ls}

\affiliation{Institute for Astronomy, University of Hawaii,
2680 Woodlawn Dr., Honolulu, HI 96822, USA}

\begin{document}
\ifnfssone
\else
  \ifnfsstwo
  \else
    \ifoldfss
      \let\mathcal\cal
      \let\mathrm\rm
      \let\mathsf\sf
    \fi
  \fi
\fi

\maketitle

\begin{abstract}

In the inflationary scenario all the matter constituting the universe
was created from the process of reheating
after inflation.
Recent development of
the  theory of reheating  is briefly
reviewed.
 The list of topics includes elementary
(perturbative) theory of reheating;
quantum field theory in a time-varying  background;
parametric resonance and explosive particle creation;
 non-thermal phase transitions from reheating;
baryogenesis from reheating;
residual oscillations of the scalar field, and other
cosmological applications.

\end{abstract}

\firstsection 
\section{Introduction}

As a graduate student, I studied   particle creation
from strong gravitational and electric fields. I was impressed
by a  paper which  Ya.~Zel'dovich (1972) devoted to J. Wheeler's
60th birthday volume.
In particular, Zel'dovich quoted there the ``Novikov paradox'':
 let an electron-positron pair be created from the
strong electric field. The total momentum of the pair must be equal
to zero, but the total energy is not zero. Query:
how can this result be the Lorentz-invariant?
Answer: the result  indeed is the Lorentz-invariant, if the space-time
interval between  electron and positron is space-like. Hence,
at the classical level, the creation of the pair is a causally independent
event.
The ``Novikov paradox'' is a fair example of how Igor  treats
 complex physical problems such as quantum   creations of particles.
 Therefore taking occasion to
 contribute to  Igor Novikov's 60th birthday  volume,
 I  would like to consider  another, one of the most
spectacular, applications of the quantum theory of particle creation
that emerges together with the  Inflationary Scenario.
Indeed,  almost all matter constituting the
Universe at the subsequent radiation-dominated stage was
created from the reheating after inflation.
200z
The term ``reheating'' here is an anachronism after the first inflationary
models in which  the
Universe was hot before inflation and  was reheated again after
inflation. In modern versions of inflationary cosmology  the
 pre-inflationary hot stage is no longer necessary  (Linde 1990).
It is assumed that the Universe initially expands quasi-exponentially
in a vacuum-like state with a  vanishing entropy and particle
number density.
At the stage of inflation, all energy was
concentrated in a classical slowly moving inflaton field $\phi$. Soon
after the end of inflation, when an observable universe was the size of
a dime, the inflaton
  field began to oscillate near the minimum of
its effective potential $V(\phi)$. An
 almost homogeneous inflaton field $\phi(t)$   coherently
oscillated with a very large  amplitude of the order of the Planck mass
 $\phi \sim M_p$.
The interaction of the inflaton field with other elementary
particles  led to creation of many ultra-relativistic
particles from the  classical inflaton oscillations.
 Gradually, the inflaton field decayed and transferred
all its energy  non-adiabatically  to the created
particles.
They interacted with each other and came to a state of thermal equilibrium
at some temperature $T_r$, which was called the reheating temperature.
An enormous
 total entropy of the observable universe (or the total number of particles
inside the horizon),  $S \sim  10^{88}$,  was thus  produced from the
reheating after inflation.

The reheating is an intermediate stage between the inflation and the
radiation dominated Friedmann expansion. Therefore
the reheating is associated with
a number of issues of the Big Bang scenario:
 the large entropy problem, the baryogenesis problem,
the  problem of relic monopoles, the problem of topological defects,
the primordial black holes problem, etc.

The elementary theory  of reheating based on the perturbation theory was
developed right after the first models of inflation were
suggested. Various aspects of the
theory of reheating  were further elaborated by many authors.
The elementary theory of reheating found its place
into the textbooks on the inflation (Linde 1990; Kolb \& Turner 1990).
 Still, the general scenario of reheating in inflationary cosmology
was absent,  and a number of important problems remained unresolved.
 In particular, reheating in  chaotic inflation theory,
which covers many popular models of inflation,
remained almost unexplored.
Recently, the  new effects  in  particle creation, arising beyond of the
perturbation theory, were found, together with  further
constraints on the elementary  theory itself.
These effects  significantly alter the reheating scenarios.
 In this contribution   we briefly review   the theory of the
reheating after inflation. 
We will discuss the  different approaches to the theory
of reheating: from a quantitative, oversimplified point of view, and within
rigorous qualitative formalism. This presentation is based on the ongoing
collaboration with Andrei Linde and Alexei Starobinsky.

\section{Evolution of the Inflaton Field}

We consider a simple chaotic inflation.
Among all the fields $(\phi, \chi, \psi, A_i, h_{ik}, ...)$
which are present at these energy scales, the main contribution to
gravity comes from the condensate of the scalar field $\phi$,
or from other condensates which effectively  similar to that.
In the fundamental
Lagrangian ${\cal L}(\phi, \chi, \psi, A_i, h_{ik}, ...)$
for large energy density,
 we can retain for the cosmological applications only gravity and
 the dominant scalar field
\begin{equation}
{\cal L}(R, \phi) = -{M_p^2 \over 16\upi}R + {1 \over 2} \phi_i \phi^i
 -V(\phi).
  \label{grav}
\end{equation}
The evolution of the FRW universe is described by the Einstein equation
\begin{equation}
H^2={{ 8\pi} \over {3 M_p^2}}\biggl( {1 \over 2}\dot \phi^2 +
V( \phi) \biggr)\ ,
  \label{E}
\end{equation}
where $H={\dot a / a}$.
The Klein-Gordon equation for $\phi(t)$
 is
\begin{equation}
\ddot \phi + 3H \dot \phi + V_{, \phi}=0\ .
  \label{KG}
\end{equation}
For a sufficiently large initial values of $\phi > M_p$,
a ``drag'' term $3H \dot \phi$  in (\ref{KG})
dominates over $\ddot \phi$;
the potential term in (\ref{E}) dominates over the kinetic term.
This is the inflationary stage, where the universe expands
 (quasi-)exponentially, $a(t)=a_0 \exp \bigl( \int dt H(t) \bigr)$.
However, with a decrease of the field $\phi$ the ``drag'' term becomes less
and less important, and inflation terminates at $\phi {\
\lower-1.2pt\vbox{\hbox{\rlap{$<$}\lower5pt\vbox{\hbox{$\sim$}}}}\ }M_p/2$.
 After a short stage
of the fast rolling down, the inflaton field
  rapidly oscillates around the minimum of $V(\phi)$
with the initial amplitude $\phi_0 \sim 0.1 M_p$.
Although this value is below the magnitude needed for inflation,
it is still a very large figure.

The character of the classical oscillations of the homogeneous scalar field
depends on the shape of its potential $V(\phi)$ around the minimum.
We will consider two models:    the quadratic
potential $V(\phi)={1 \over 2} m_{\phi} \phi^2$ and
  the  potential
 $V(\phi)={1 \over 4} \lambda \phi^4$.

For the quadratic potential the solutions of eqs. (\ref{E}) and  (\ref{KG})
at the stage of   oscillations are
\begin{equation}
\phi(t) \approx \phi_0(t) \cdot
 \sin{\left( m_{\phi}t \right)}\ , \ \ \
 \phi_0(t)= {M_p \over \sqrt{3\pi}}\cdot{1  \over m_{\phi}t}\ .
\label{87}
\end{equation}
The scalar factor averaged over several oscillations is
$ a(t) \approx a_0 t^{2/3}$.
Oscillations of $\phi$ in this theory are sinusoidal, with the
 the amplitude
$\phi_0(t) \sim {0.1 M_p } {a^{- 3/2}}$
 decreasing as the universe expands.
The energy
of the field $\phi$ decreases in the same way as the density of
nonrelativistic
particles of mass $ m_{\phi}$:
 $\epsilon_{\phi} = {1\over2} \dot \phi^2  +
  { 1\over2 } m_{\phi}^2 \phi^2 \sim a^{ -  3}$.
Hence the coherent oscillations of the homogeneous scalar field
correspond to the
  the matter dominated effective equation of states with vanishing
pressure.

For the theory with the potential  $V(\phi)={1 \over 4} \lambda \phi^4$
 it is more convenient to express
 the solutions of  eqs. (\ref{E}) and  (\ref{KG})    via the conformal time
  $\eta = \int {dt\over a(t)}$:
\begin{equation}
\phi(\eta) \approx \phi_0(\eta) \cdot
  cn \left({\omega \over c} \eta, { 1 \over \sqrt{2}}\right)
  \ , \ \ \  \phi_0(\eta)=
 { \sqrt{3 \over 2\pi}}M_p \cdot
 { c \over {\omega \eta}} \ .
\label{87a}
\end{equation}
The oscillations in this theory  are not sinusoidal, but given by elliptic
 function\hfil\break
  $cn({\omega \over c} \eta, { 1 \over \sqrt{2}})$,
 with the period of oscillations
$ {2\pi \over  \omega}$, where a numerical constant $c \approx 0.85$ and
the effective frequency of oscillations
 $\omega=c \sqrt{\lambda}a\phi_0  \sim 0.1 \sqrt{\lambda} M_p$.
With a good accuracy one can write
$\phi(\eta)
\approx \phi_0(\eta) \sin (c\sqrt \lambda  a \phi_0 \eta)$.
The scalar factor is $a(t) \propto \eta(t) \propto \sqrt{t}$.
 The amplitude   of oscillations
$\phi_0(t) \sim  {0.1 M_p } {a^{- 1}}$
 decreasing as the universe expands.
The energy
of the field $\phi$ decreases in the same way as the density of
relativistic particles:
 $\epsilon_{\phi} = {1\over2} \dot \phi^2  +
  { 1\over 4 } \lambda \phi^4 \sim a^{ -  4}$.
The effective equation of state
corresponds to
  the radiation dominated  equation of states with
pressure $p_{\phi} \approx {1 \over 3} \epsilon_{\phi}$.

\section{Elementary Theory of Reheating }

 The amplitude of the oscillations  gradually
decreases not only because of expansion of the
universe, but also because of the energy transfer to particles
created by the
oscillating field.
 To describe  the decay of the inflaton oscillations,
we shall remember again the rest of the fundamental Lagrangian
 ${\cal L}(\phi, \chi, \psi, A_i, h_{ik}, ...)$ which
includes the other fields and their interaction with the
inflaton field.
 The inflaton field $\phi$
after inflation may decay
into bosons $\chi$ and fermions $\psi$ due to the interaction  terms $- {
1\over2} g^2 \phi^2 \chi^2$ and
 $- h \bar \psi \psi \phi$, or into its own Bose quanta $\delta \phi$
due to the self-interaction $\lambda \phi^2 \delta \phi^2$,
or due to the gravitational interaction (like in
the Starobinsky  inflationary model).
 Here $\lambda$, $ g$ and  $h$ are
small coupling constants.  In case of  spontaneous symmetry breaking, the term
$- {
1\over2} g^2 \phi^2 \chi^2$ gives rise to the   term  $- g^2 \sigma\phi
\chi^2$.
We will assume for simplicity that the bare masses
of the fields $\chi$ and $\psi$ are very small, so that one can write  $ m_\chi
(\phi) \approx
  g \phi$,  $m_{\psi}(\phi) \approx  |h\phi|$.

An elementary theory  of reheating based on the perturbation theory
 was developed by Dolgov \& Linde (1982) and by
Abbot, Fahri \& Wise (1982) for the new inflationary scenario.
At the same time
the theory of reheating was constructed in the Starobinsky (1982) model.
Let us briefly recall the elementary theory of reheating.
We consider for simplicity  the
classical  field $\phi (t)$  with
 the mass $ m_{\phi}$  oscillating near the minimum
 of the quadratic potential.
 A homogeneous scalar
field oscillating with  frequency $\omega  = m_{\phi}$
  can be interpreted
 as a
collection of a number of $\phi$-particles with zero
momenta.
The coherent wave of  particles at rest
with  energy $\epsilon_{\phi}$
 corresponds to particle density $n_{\phi} = \epsilon_{\phi}/m_{\phi}$.
 Another way around, $n_{\phi}$  oscillators of the
same
frequency $m_{\phi}$, oscillating coherently with the same
phase, can be viewed as a single homogeneous wave $\phi(t)$.
Based on that interpretation,
the  effects related to the
 particle production can be incorporated into
 the equation of motion (\ref{KG}) for the  inflaton field
by means of the polarization operator
(Linde 1990):
\begin{equation}\label{2}
\ddot \phi  +   3 H \dot \phi  +   \left( m_{\phi}^2  +    \Pi
(\omega) \right) \phi  = 0\ .
\end{equation}
Here $\Pi (\omega)$ is the flat space polarization operator for the
field $\phi$ with the four-momentum $ k_i  = (\omega , 0,0,0),\, \omega =
m_{\phi}$. The real part of $\Pi(\omega)$ gives only a small correction to
 $m_{\phi}^2$,  but when $ \omega \ge min(2m_{\chi},2m_{\psi})$, the
polarization operator $\Pi(\omega)$  acquires an imaginary part $
\mbox{Im}\,
\Pi(\omega)$. We will assume that
 $m_{\phi}^2 \gg H^2, m_{\phi}^2 \gg \mbox{Im}\, \Pi$. The first
condition
is
automatically satisfied  after the end of inflation; the second
usually is also
true.
The
 solution of (\ref{2}) that generalizes the solution (\ref{87})
and
describes
damped oscillations of the inflaton field is
\begin{equation}
\phi    \approx {M_p \over {\sqrt{3\pi} m_{\phi}t}}
 \exp {\left(  -
{1\over2} \Gamma t \right)} \sin{\left( m_{\phi}t \right)}\ ,
\label{3}
\end{equation}
where  $\Gamma$ is the total decay rate of $\phi$-particles. Here we used
a relation
${\mbox{Im}}\, \Pi  = m_{\phi} \cdot \Gamma $,
 which follows from unitarity.
Thus, eq. (\ref{3}) implies that the amplitude  of oscillations of
the field $\phi$ decreases as
$\phi_0(t) \approx  {0.1 M_p} { a^{-3/2}}
\exp {\big (  -  {1\over2}
   \Gamma  t \big)}$ due to particle production which
occurs during  the decay of the inflaton field, as well as due to
the expansion
of the universe.

For a phenomenological
description of this effect one can just add an extra friction term
$\Gamma\dot\phi$ to the classical equation of motion of the field
$\phi$,
instead of adding the
polarization operator (Kolb \& Turner 1990):
\begin{equation}
\ddot\phi  +   3 H \dot\phi  +    \Gamma \dot\phi  +
m_{\phi}^2\phi  = 0\ .
\label{5}
\end{equation}
At the stage of oscillations the solution of this equation
is similar to  (\ref{3}).

One should note, that this equation, just as  equation (\ref{3}), is
valid only under the conditions $m_\phi \gg H$, $m_\phi \gg \Gamma$,
and only at the stage of rapid oscillations of the field $\phi$ near
the minimum of $V(\phi)$. This equation  cannot be used to
investigate of the stage of
 slow and fast rolling of the field $\phi$ during
inflation.

Suppose that $\Gamma^{-1} $ is  much less than
the
typical expansion time of the Universe $ H^{ -  1}$,
 the energy density of the inflaton field decreases
exponentially within the time $\Gamma^{-1}$:
$\epsilon_{\phi} \approx
 {1 \over 2} m_{\phi}^2
\phi_0^2 \cdot \exp( -  \Gamma t) .$
This is exactly the result one would expect on the basis of the
interpretation
of the oscillating field  $\phi$  as a coherent wave consisting of
decaying
$\phi$-particles.

The rate of decrease of
the energy of oscillations
coincides with the decay rate  of  $\phi$-particles
which is well-known from the perturbation theory.
 The
rates of
the processes $\phi \to \chi\chi$  and $\phi \to  \psi\psi$ (for  $m_\phi \gg
2m_\chi, 2m_\psi$) plus gravitational decay
 are given
by
 \begin{equation}\label{7}
  \Gamma_{ \phi \to \chi \chi} =  { g^4 \sigma^2\over 8
\pi m_{\phi}}\  , \ \
\Gamma_{ \phi \to \psi \psi }  =  { h^2 m_{\phi}\over 8 \pi}\ , \ \
\Gamma_g = {m_\phi^3 \over 8\pi M_p^2}\ .
 \end{equation}
Reheating
completes at the moment $t_r$
 when the rate of expansion of the universe given   by the Hubble
constant $H=\sqrt{8\pi \rho\over 3 M^2_p} \sim  t_r^{-1}$ becomes equal to
the total decay rate $\Gamma$. The energy density of the universe at the
 moment  $t_r=\Gamma^{-1} $ is
\begin{equation}
\epsilon(t_r)
\simeq {3 \Gamma^2M_p^2 \over 32 \pi}\ .
\label{en1}
\end{equation}
 If the
particles produced by the decaying inflaton field interact with each
other
strongly enough, then
the thermodynamic equilibrium sets in quickly after the
decay of the inflaton field, and the matter acquires a temperature
$T_r$.
 The energy
density of the Universe  dominated by the ultrarelativistic
particles in a
state of thermal equilibrium is
\begin{equation}
\epsilon(T_r)
\simeq   {\pi^2 g_* \over 30}T_r^4\ ,
\label{en2}
\end{equation}
where  a numerical factor $g_*(T_r) \sim 10^2 -  10^3$ depends on
the number of ultra-relativistic  degrees of freedom.
Comparing (\ref{en1}) and  (\ref{en2}) we get the
following result for the reheating temperature:
\begin{equation}
T_r \simeq 0.1 \sqrt{\Gamma M_p}\ .
\label{12}
\end{equation}
Note that $T_r$ does not depend on the initial value
of the field  $\phi$, it is completely
determined by the parameters of the underlying elementary particle
theory via the decay rate $\Gamma$.

In order to get numerical estimates for the duration of reheating
$t_r$, for the decay rate of the
inflaton field and for the reheating temperature $T_r$, one should
know the mass  of the inflaton field
 and the coupling constants.
We shall take into account
 several constraints on these
in the framework of the perturbation theory
(Kofman, Linde \& Starobinsky 1996b).
 The coupling constants of interaction of
the inflaton field with matter cannot be too large, otherwise the
radiative corrections  alter the shape of the inflaton potential.
The necessary condition for the decay of inflaton oscillations is
$\omega > 2m_{\chi}, 2m_{\psi}$.
Parameters of the inflaton potential are restricted from the constraints
on amplitude of the cosmological fluctuations.
All together it allows us to update a constraint on the elementary theory:
 the largest possible total decay rate
in the perturbation theory is
$\Gamma < 10^{ -  20} M_p $.
For instance, for the quadratic inflaton potential, it takes
at least
$10^{14}$ oscillations  to
transfer the energy of inflaton oscillations into
  the created particles.
This is a very strong condition, which makes reheating very slow.
 From (\ref{12})
 one can
obtain the  general bound on the reheating temperature in
the model of slow reheating:
\begin{equation}
 T_r <   10^9 \, \mbox{GeV} \ .
\label{T}
\end{equation}
This is a very small temperature, at which the standard mechanism of
baryogenesis in the GUTs cannot work; in this case
 we definitely need
a theory of a low-temperature baryogenesis.
At such a
temperature
 no cosmologically
interesting
heavy  strings, monopoles and textures can be produced, because the
generation of the  GUT
topological  defects require  the high temperature phase transitions
at $T_{GUT} \sim 10^{16}  \, \mbox{GeV} \ .$
For the massive inflaton field,
in the slow reheating scenario
the post-inflationary    matter dominated equation of state of the
inflaton oscillations   lasts sufficiently long. It can enhance
the gravitational instability of density fluctuations,
which  leads to the formation of the primordial black holes
for some  specific spectra of the initial density fluctuations.

\section {Quantum Field Theory in a Time-dependent Background}

Surprisingly,
it was found recently that  typically the transition between inflation and the
hot Big Bang universe may be very different from what the elementary theory
predicts (Kofman, Linde \& Starobinsky 1994;
Shtanov,  Traschen \&  Brandenberger 1995; Boyanovsky {\it et al} 1995a;
Yoshimura 1995).
The elementary theory of reheating might be still
applicable  if the inflaton field
can decay  into fermions only, with a small coupling constant $h^2 \ll
m_{\phi}/M_p$.
This theory also can provide a qualitatively correct
description of particle decay at the last stages of reheating.
 However,
 the perturbation
 theory is inapplicable to the description of the first stages of
reheating, which makes the whole process quite different. In what follows we
will primarily
consider the theory of the first stages of reheating. We will begin with
the theory of a massive scalar field $\phi$ decaying into particles $\chi$,
then we consider the theory  ${\lambda\over 4} \phi^4$.

For simplicity, we
  consider here the interaction $-{1 \over 2}g^2 F(\phi) \chi^2$ between
 the {\it classical }inflaton field $\phi$
and the {\it quantum}  scalar field
$\chi$ with the Lagrangian:
\begin{equation}
{\cal L}( \chi) = {1\over2} {\left( \chi_{,k}\chi^{,k}  -  m_{\chi}^2 \chi^2  +
\xi R \chi^2  -  g^2 F(\phi) \chi^2 \right)}\ .
\label{int}
\end{equation}
Here  $\xi$ is a coupling constant of interaction with
  the space-time curvature $R$. We also will distinguish
two  cases, the four-legs interaction
$g^2 F(\phi)\chi^2=g^2 \phi^2 \chi^2$, and the three-legs
interaction $g^2F(\phi)\chi^2=2g^2\sigma \phi\chi^2$ which arises
when symmetry is broken.
 The Heisenberg representation of the quantum scalar field $\hat \chi$ is
\begin{equation}
\hat \chi(t, \vec x)  =
{1\over{(2\pi)^{3/2}}} \int d^3k\ {\Bigl( \hat a_{\vec k} \chi_{\vec
k}(t)\, e^{ -
i{\vec k}{\vec x}}
+    \hat a_{\vec k}^ +    \chi_{\vec k}^*(t)\, e^{i{\vec k}{\vec x}}
\Bigr)}\ ,
\label{37}
\end{equation}
where $\hat a_{\vec k}$ and $\hat a_{\vec k}^ +   $ are the
annihilation and creation
Bose-operators. For the flat Friedmann background with a scalar
factor $a(t)$ the temporal
part of the eigenfunction with momentum $\vec k$  obeys the equation
\begin{equation}
\ddot \chi_k  +    3{{\dot a}\over a}\dot \chi_k +   {\left(
{k^2\over a^2}
 +   m^2_{\chi} -  \xi R  +  g^2F(\phi) \right)} \chi_k  = 0 \ .
\label{38}
\end{equation}
Let us use the conformal time $\eta$ and introduce the new
function $f_k(\eta) = a(\eta)\chi_k(\eta)$. Then instead of
(\ref{38}) we have
\begin{equation}
f_k''  +    \Omega_k^2f_k  = 0\ ,
\label{39}
\end{equation}
where prime stands for ${d\over{d\eta}}$, and
$ \Omega_k^2 = (m_{\chi}a)^2  +    k^2  +     g^2 a^2 F(\phi)
+a^2(1/6- \xi)R$.
This equation describes the  oscillators with a variable
frequency $\Omega_k^2(t)$ due to the time-dependence of the
background field $\phi(\eta)$ and $a(\eta)$.
At $\eta \to - \infty$
we choose  the  positive-frequency solution
$ f_k(\eta) \simeq { {e^{-ik\eta}}\over
\sqrt{2k}}$.
We expect the quantum effect of the $\chi$-particles creation and
 vacuum polarization as the inflaton field is varying during
the slow- and fast-rolling down  and oscillations
after inflation.
 The  problem is to find a
general solution of the  equation (\ref{39})
 and regularize the formal
expressions for the  values of the energy-density
$ \langle \epsilon_{\chi} \rangle$ and
the vacuum expectation  $ \langle \chi^2 \rangle$.

We adopt a physically transparent method to treat the
eq.(\ref{39})
 for an arbitrary time dependence of the classical
background field which was originally developed by
 Zeldovich and Starobinsky (1972)  for the problem of the
 particle creation in  varying strong gravitational field.
In terms of classical waves of the $\chi$-field,
quantum effects occur due to the departure from the initial
positive-frequency solution.
Therefore one
may  represent solutions of eq.(\ref{39}) as a product of its
solutions in the  adiabatic approximation,  $\exp{( \pm  i\int d\eta\
\Omega_k)}$, multiplied by some functions  $\alpha(\eta)$ and
$\beta(\eta)$:
\begin{equation}
 f_k(\eta) =
{\alpha_k(\eta)\over \sqrt{2\Omega_k}}\ e^{-  i\int
\Omega_k  d\eta}
 +   {\beta_k(\eta)\over \sqrt {2\Omega_k}}\ e^{+   i\int \Omega_k
d\eta} \ .
\label{61}
\end{equation}
An additional condition on the functions $\alpha$
and $\beta$ can be imposed by taking the derivative
 of the expression (\ref{61})
as if  $\alpha$ and $\beta$ would be time-independent.
Then the expression (\ref{61}) is a solution of equation
(\ref{39})  if the functions
$\alpha_k, \beta_k$ satisfy the equations
\begin{equation}\label{63}
\alpha_k' = {\Omega_k' \over 2\Omega_k} e^{+ 2  i\int \Omega_k
d\eta}\ \beta_k\ ,~~\  \
\beta_k' = {\Omega_k' \over 2\Omega_k}  e^{-  2 i\int
\Omega_k d\eta}\ \alpha_k \ .
\end{equation}
Normalization gives
 $\vert \alpha_k \vert^2 -  \vert \beta_k \vert^2 = 1$,
 and the initial conditions  at   $\eta \to  - \infty$ are
$\alpha_k = 1$, $\beta_k  =  0$.
 The coefficients $\alpha_k(\eta)$ and
$\beta_k(\eta)$ in our case coincide with the coefficients of the
Bogoliubov
transformation of the creation and annihilation operators,
 which
diagonalizes the Hamiltonian  of the $\hat \chi$-field
 at each moment of time $\eta$.

The regularized vacuum expectation values for the
$\chi^2$, energy and particle number densities
in terms of $\alpha_k(\eta)$ and $\beta_k(\eta)$ are given
correspondingly
by
\begin{equation}
\langle \chi^2 \rangle
 =  {1 \over 2\pi^2 a^2} \int_0^{\infty} dk\, k^2 {1 \over
\Omega_k} {\left(    \vert \beta_k \vert^2  +     \mbox{Re} \,
(\alpha_k\, \beta_k^*\,  e^{-2i\int
\Omega_kd\eta}) \right)_{reg}}\ ,
\label{65}
\end{equation}
\begin{equation}
\langle\epsilon_{\chi}\rangle  = {1 \over 2\pi^2 a^4}
\Bigl(    \int_0^{\infty}
dk\, k^2 \Omega_k\   \vert \beta_k \vert^2 \Bigr)_{reg}\ , \ \
\langle n_{\chi}\rangle  = {1 \over 2\pi^2 a^4}
 \int_0^{\infty}
dk\, k^2  \vert \beta_k \vert^2\
{ }.
\label{number}
\end{equation}
The formal expressions for the vacuum expectation values
 should  be renormalized. The WKB expansion of the solution of eqs. (\ref{63})
provides a natural scheme of regularization (Zel'dovich \& Starobinsky 1972).
 Thus, the final result is expressed via the coefficients of the
Bogoliubov transformation.

  The equation
of motion of the inflaton  field with the feedback of the quantum effects is
\begin{equation}
\ddot \phi   +    3H \dot \phi   +     V_{,\phi}  +
{1 \over 2} g^2 \langle \chi^2\rangle F_{, \phi} = 0\ ,
\label{35}
\end{equation}
 where the vacuum average is given by eq.(\ref{65}).
The quantum effects contribute to the effective mass of the
 inflaton field,  e.g., for the massive scalar field
$m^2_{eff}=m^2_{\phi} + {1 \over 2}g^2 \langle \chi^2 \rangle F_{,\phi}$.
The interactions with fermions
  $ -h \bar \psi \psi \phi$ and the self-interaction
$-\lambda \phi^2 \delta \phi^2$, where $\delta \phi = \hat\phi - \phi$,
result in extra terms
 $ h<\bar \psi \psi>$ and
$3\lambda \phi^2  \langle \delta \phi^2 \rangle$
in (\ref{35}).

With the self-consistent equations (\ref{65}) and (\ref{35})
based on the
rigorous quantum field theory in the time varying
background, one can investigate the role of the quantum effects
on different stages of the evolution of the inflaton field.
There are three main  quantum effects  in an external classical
background: vacuum polarization, particle creation and generation
of long-wavelength semi-classical fluctuations.
Different effects dominate at different stages of the inflationary
scenario.
During inflation
the  long-wavelength fluctuations are generated
  for the minimally coupled light fields with  $\xi=0$,
 but this effect is irrelevant to the reheating.
 The vacuum polarization and particle production could occur
at the next stage of the fast rolling down
 of the inflaton field
to the minimum of its potential $V(\phi)$ at the end of inflation.
 However, these effects are negligible
and do not alter the dynamics of the inflaton field.
At the stage of oscillations of the inflaton field
 the only important effect is the creation of particles due to the
non-adiabatic change of $\Omega_k(\eta)$.
 The particle production leads
to the generation of $\langle \chi^2 \rangle, ~
\langle \epsilon_{\chi} \rangle$ and
 the damping of the inflaton oscillations.

However, the term $ {1 \over 2}g^2\langle \chi^2 \rangle F_{,\phi}$
 in the equation (\ref{35})
 which describes the
back reaction of the quantum effects on the evolution of the
inflaton field is not reduced to the phenomenological ``friction'' term
  $\Gamma\dot\phi$, first suggested by  Albrecht {\it et al} (1982).
This oversimplified
ansatz is  widely  used to investigate
all the stages of the evolution of the  inflaton field.
Many authors
used this method in order
to understand whether reheating may slow down the rolling of the
field $\phi$  and,
consequently, to support inflation.
 Remember
that one
of the main reasons  why
inflation is possible is the presence of the friction term
$3H\dot\phi$ in the
equation of motion of the scalar field in an expanding Universe. The
presence of a similar term due to reheating could support inflation
even in the models where inflation would otherwise  be
impossible.
As we seen, the  damping of
 the scalar field during slow and fast rolling down
 cannot be described in
such a simple way. In our study (Kofman, Linde \& Starobinsky 1996b)
we did not find any
possibility to support inflation by reheating.
As an exception, the friction  term  can correctly
describe  damping of oscillations
in the case where  the perturbation theory for
interaction is applicable.

It is instructive
 to return  in the framework of the
rigorous  approach
 to the perturbation theory. Assuming  small  $|\beta_k| \ll 1$,
  from eqs. (\ref{63}) one can
obtain an iterative solution:
\begin{equation}
\beta_k \simeq
{1 \over2} \int_{ -  \infty}^{\eta} d\eta'\,{\Omega_k'\over
\Omega_k}\, \exp{\bigl( -
 2i \int_{ - \infty}^{\eta'}d\eta'' \Omega_k(\eta'')\bigr)}\ .
\label{BET}
\end{equation}
Using expression  for $\Omega_k(\eta)$, we obtain that
$\beta_k(\eta)$ is proportional to
 the integral \hfil\break
$ g^2  \int_{ -\infty}^{\eta}d\eta'\,  e^{-2ik\eta'}\, (a^2 F(\phi))'$,
where $\phi(t)$ is given by the oscillating
 solutions (\ref{87}) or (\ref{87a}), and
$F(\phi)$ depends on the type of the interaction.
 This integral is evaluated
by the method of stationary phase (e.g., Starobinsky 1982).
To compare the results with these derived in the previous section,
let us focus on the case of massive scalar field
decaying via  the  three-legs interaction $g^2 \sigma \phi \chi^2$.
In this case the dominant contribution  is given by the
integration near  $\eta_k$, where $a(\eta_k)={2k \over m_{\phi}}$.
This corresponds to the creation of
a pair of  $\chi$-particles with  momentum $k ={ 1\over 2} a(\eta_k)m_{\phi}$
 from an  inflaton with the four-momentum
$(m_{\phi}, 0, 0, 0)$ at the moment $\eta_k$.

Substituting this  estimation
for $\beta_k$ into (\ref{number}),
it is easy to calculate the decay rate of the inflaton field.
For the three-legs
interactions $g \sigma \phi \chi$ or $h \phi \bar \psi \psi$ it
is indeed reduced to the formulas (\ref{7}).
For  the four-legs
 interaction $\phi \phi \to \chi \chi$, we have creation of
a pair of  $\chi$-particles with  momentum $k = a(\eta_k)m_{\phi}$
 from a pair of massive inflatons with the four-momentum
$(m_{\phi}, 0, 0, 0)$ at the moment $\eta_k$.
The  decay rate of the massive inflaton field in this case
is rapidly decreases with the expansion of the Universe as
${1\over a^4}{d \over dt}(a^4
\epsilon_{\chi}) \propto a^{-6}.$
 Therefore the complete
decay of the massive inflaton field  in the theory with
 no spontaneous symmetry breaking or with  no interactions with fermions
 is impossible! This  observation
 gives rise to an unexpected possibility to consider the residual
 scalar field oscillations as a  dark matter candidate.
Indeed, if the inflaton oscillations correspond to the matter
dominated equation of state,
the inflaton field  by
itself, or other scalar fields with similar properties can be
cold dark matter candidates,  even if they strongly interact with each
other. However, this possibility requires
a  fine tuning. More immediate application of our result is
that it allows one to rule out the inflationary models which
predict too small or too large
value of the present  radiation density parameter $\Omega_r$.

The fact that  the number
of  particles  $|\beta_k(\eta)|^2$
 with the momentum $k$ is predominantly generated
at the certain moment $\eta_k$
gives us a proper physical interpretation how the particles are created
from the coherent oscillations of the homogeneous inflaton field:
 a pair of particles
is created at the instant of the resonance
 $k=\omega \cdot a(\eta)/2$ between $k$ and the external
 frequency $\omega$ (which is equal $m_{\phi}$ for the massive inflaton).
 As the universe expands,
that particular mode is redshifted from the resonance,
and creation of particles with the momentum $k$ is terminated
leaving $ n_k=|\beta_k(\eta)|^2 \ll 1$.
At an arbitrary moment $\eta$ there is a corresponding
momentum $k$ resonating with the external frequency.

\section{Parametric Resonance: the Stage of Preheating }

A new interpretation
 that particles are created at the instant of resonance
with  the inflaton oscillations
 we have grounded in the last section
 gives us a hint for the
 general case when   $n_k$ is not assumed to be
small.

Let us  assume there is no symmetry breaking prior reheating.
Therefore  we  will consider oscillations (\ref{87}) of the
  massive scalar field $\phi$  decaying into light particles
 $\chi$
 due to the interaction $-{ 1\over2} g^2 \phi^2  \chi^2$;
and oscillations (\ref{87a}) in the theory $\lambda \phi^4$
decaying due to the self-interaction
 $-{ 3\over 2} \lambda \phi^2 \delta \phi^2$.
In the first case
the equation (\ref{38}) for quantum fluctuations  $\chi_k(t)$
can be rewritten in the   form
\begin{equation}
{d^2 (a^{3/2}\chi_k) \over dt^2}   +   \left({k^2\over a^2(t)} +
+ g^2 \phi_0^2\, \sin^2(m_{\phi}t) + \Delta \right)(a^{3/2} \chi_k) = 0 \ ,
\label{M}
\end{equation}
where  $\phi_0$ stands for the amplitude of
oscillations of the field $\phi$, defined in the formula (\ref{87}),
$\Delta = m_{\chi}^2 + (9/2 -3\xi)H^2$.
 As we shall see, the
main contribution to $\chi$-particle
production is given by excitations of the field $\chi$ with
 $k/a > m_\phi$, which is much
greater than $H$ at the stage
of oscillations. Therefore, in the first approximation we may neglect
the expansion of the Universe,  taking $a(t)$ as a constant and omitting
the term $\Delta$ in (\ref{M}).

In the case of theory with the   ${\lambda \over 4} \phi^4$ potential
the role of quantum field  plays the quantum fluctuations
$\delta \phi$. In the equations  (\ref{38}) for the quantum fluctuations
we have to substitute $\chi_k \to \delta \phi_k$,
$g^2 \to 3 \lambda$. Since in this theory the amplitude
$\phi_0 (t) \propto a^{-1}(t)$, it is convenient to
use the  conformal time $\eta$ and equations for fluctuations in the
 form (\ref{39}). For the oscillations (\ref{87a}) an
 exact equation for quantum fluctuations
 $\delta \phi$ can be reduced to the Lame equation.
For simplicity here we will use an  approximate equation
\begin{equation}
{d^2(a \delta\phi_k)\over d\eta^2}   +   {\Bigl[{k^2} +
3\lambda a^2\phi_0^2\, \sin^2 (c\sqrt\lambda a\phi_0 \eta)\Bigr]}
(a \delta\phi_k) = 0 \ ,
\label{lam1}
\end{equation}
where  $c \approx 0.85$,
see (\ref{87a}). The crucial observation is that the equations
 (\ref{M}) and (\ref{lam1}), which
 describe the fluctuations in the models, can be reduced to
 the well-known  Mathieu equation:
\begin{equation}
{{ d^2 y_k} \over dz^2}  +   \left(A(k) - 2q \cos 2z \right) y_k = 0 \ ,
\label{M1}
\end{equation}
with the periodic effective frequency $\Omega_k^2=A(k) - 2q \cos 2z$.
For the first model (\ref{M}),  $y_k=a^{3/2}\chi_k$,
  $A(k)
= {k^2 \over \omega^2 a^2}+2q$, $q = {g^2\Phi^2\over
4\omega^2} $, $z
= \omega t$, the frequency of the inflaton oscillations $\omega=m_{\phi}$.
For the second model (\ref{lam1}),  $y_k=a \cdot \delta \phi_k$,
 $A \approx  {k^2\over \omega^2} + 2.08$,
 and
$q  \approx 1.04$, $z=\omega \eta$, the frequency of inflaton oscillations
$\omega =c \sqrt{\lambda} a\phi_0$.

An important property of solutions of the equation (\ref{M1}) is the
existence of an exponential instability $y_k \propto \exp
(\mu_k^{(n)}z)$ within the set of resonance bands  of frequencies
$\Delta k^{(n)}$ labeled by an integer index $n$.
This instability corresponds to exponential growth of occupation
numbers of quantum fluctuations
$n_{\vec k}(t) \propto \exp (2\mu_k^{(n)} m_{\phi} t)$
  that may be interpreted as particle
production. The simplest way to analyse this effect is to study the
stability/instability chart of the Mathieu equation
(MacLachlan 1961).

We will distinguish different regimes of the resonance which correspond
to the different regions of the parameters of the
instability chart:

{\bf  Perturbation theory: } $q \ll 1 \ ; \  A=l^2, \ l=1, 2, ... \ ; \
2\pi\mu_k \ll  H/\omega \ll 1 \ ; \ n_k \ll 1 \ .$
If no expansion of the universe, fluctuations $y_k$ are slowly
generated at the discrete modes $k$. The expansion quickly
 redshifts them from the resonance bands.
 The net effect is reduced to the
 creation of  $n_k \ll 1$ particles with continuous
spectrum, which we    considered in the previous section.

 From this point of view the perturbative regime for the Starobinsky model
was considered by Starobinsky (1982) and for the chaotic inflation by
Kofman, Linde \&
Starobinsky (1996b).

{\bf Narrow resonance: } $q \ll 1 \ ; \  A \simeq l^2, \ l=1, 2, ... \ ; \
2\pi\mu_k \leq H/\omega \ll 1 \ ; \  n_k \leq O(1) \ .$
 The universe expansion
redshifts the momentum $k/a$ from the narrow resonance band
not too fast, so a nonsmall number of particles is
generated at the resonance mode. This is an intermediate regime between the
perturbation theory and the regime of the broad parametric resonance
where expansion of the universe is a subdominant effect.

 The narrow resonance
was mentioned by Dolgov \& Kirilova (1990)
for the new inflationary scenario.
The importance of this regime for that model
 was   first recognized
by Traschen \& Brandenberger (1990),
 but for several reasons their
final results were not  quite correct. A detailed theory of
particle creation in the
 narrow resonance regime in an expanding universe
 was developed for the chaotic scenario
by Kofman, Linde \& Starobinsky (1994; 1996b), see also Shtanov,  Traschen
\&  Brandenberger  (1995) and
 Kaiser (1995).
In order to investigate the explosive particle production from
reheating,
many  authors    considered the  model of the
narrow resonance without taking into account expansion  of the universe
 because
it made investigation simpler (Yoshimura 1995;
 Boyanovsky {\it et al.} 1995a,b).
 However for some range of parameters
many vital features of the reheating may disappear in this
approximation. In particular, the effects  in the
model  studied  by  Son  (1996) disappear in an
expanding universe.

{\bf Broad resonance:  } $q \simeq O(1) \ ; \  A \geq 2q \ ; \
2\pi\mu_k \sim O(1) \ ; \ n_k \gg 1 \ .$
Creation of
particles in the regime of a broad  resonance
 is very different from that in  the usually
considered previous cases.
 In particular, it
proceeds during a tiny part of each oscillation of the field $\phi$
when $1-\cos z \sim q^{-1}$ and the induced effective mass of the
fluctuations $y_k$  is less than $m_{eff}$.
 As a result, the number of the Bose
particles grows exponentially fast.
This regime occurs only
if parameter  $q$ is nonsmall.
 A typical energy $E$ of a particle produced at this stage is
determined by equation $A-2q \simeq 4 \sqrt{q}$.
 The line $A = 2q$ divides the
region of the broad resonance from the
tachyonic   regime where  $ A \leq 2q \ ; \
\mu_k \geq \pi^{-1} .$
 The line $A = 2q$ also
corresponds to the momentum $k = 0$.
 Near the line $A = 2q$  typically
 $\mu_k \sim  0.175$ in the instability bands,
with the maximal value about $ 0.28$.

This regime is  the    most difficult to study.
The investigation can be advanced with  the following method:
consider eq. (\ref{M1}) as the equation for the propagation of the
wave $y_k(z)$ backward in time. The evolution of  $y_k(z)$ can be
  estimated for
each scattering instance, then the net effect is the
product of the successive scattering matrixes
(Kofman, Linde \& Starobinsky 1996b).
The results for particle production based on this method
were reported by
Kofman, Linde \& Starobinsky (1994). This stage also was studied by
 Boyanovsky {\it et al.} (1995a)
and Fujisaki {\it et al.} (1995).

We shall focus on  the most interesting regime of the broad resonance.
For  our first model (\ref{M})
 the necessary condition for it
is  $m_{\phi} \ll gM_p$.
A typical energy $E$ of a particle produced at this stage is
$E  \sim  \sqrt{g m_\phi M_p}.$
As we will see,  the broad resonance (preheating)  ends up within
the short time
$t_{ph}\sim m_{\phi}^{-1} \ln (m_{\phi}/g^5M_p)$.
As a results, at the end of this stage the occupation number of
 created particles $n_k \sim exp(2 \mu_k m_{\phi}t)$
with the energy $k=E$ is extremely large:
\begin{equation}
n_E \sim {1 \over g^2} \gg 1 \ .
\label{occup1}
\end{equation}

For another model
(\ref{lam1}), the
 parameter $q =1.04$  corresponds to the broad resonance regime.
  Looking at the instability chart, we see
that the
resonance occurs in the second band.
The typical
energy $E$ of a created particles is
$E  \sim  \sqrt{\lambda} a \phi_0.$
In the second band
maximal value of the coefficient $\mu_k \approx 0.07$.
Note that the  rigorous equation Lame for the fluctuations in the
$\lambda \phi^4$ theory gives a twice smaller value of
 $\mu_k \approx 0.036$.
 As long as the backreaction of created
particles is small, expansion of the Universe does not shift fluctuations away
from the resonance band, and the
number of  particles   $n_k \sim
\exp (  {\sqrt\lambda\phi_0 \over 10}t)$.
 The broad resonance in this model ends up within
the  time interval
$t_{ph}\sim M_{\phi}^{-1}\lambda^{-1/2} |\ln \lambda|$.
At the end of this stage the occupation number is
\begin{equation}
n_E \sim {1 \over \lambda} \gg 1 \ .
\label{occup2}
\end{equation}

In addition to this models, the field $\phi$ in the theory
$\lambda \phi^4$
 may decay  to
$\chi$-particles due to the interaction $-{1 \over 2}g^2 \phi^2\chi^2$.
This is the leading process for    $g^2\gg \lambda$.
The  parametric resonance is broad. The values of the parameter
$\mu_k$
along the line $A = 2q$ do not change monotonically, but typically
for $q \sim g^2/\lambda  \gg
1$ they are 3 to 4 times greater than the parameter $\mu_k$ for the
decay of
the field $\phi$ into its own quanta. Therefore, the resonance
in this theory
 is very efficient.

The stage of broad parametric resonance which leads to the
explosive particle creation is lasting typically
for a few dozens of oscillations (for several oscillations in the
last model).
However, this does not mean that the process of reheating has been completed.
Instead of created particles in the thermal equilibrium,
 by the end of this stage
  one has particles far away of equilibrium
but with extremely large
mean occupation numbers (\ref{occup1}), (\ref{occup2}).
It is therefore
convenient to divide
 the whole process of
reheating into
  three different stages.  At the first stage, which cannot be
described by the elementary theory of reheating,  the classical coherently
oscillating
inflaton field $\phi$ decays into  bosons
 due to parametric resonance.  In many models the resonance is
 broad, and the process occurs extremely
rapidly (explosively).  Because of the Pauli exclusion principle, there is no
explosive  creation of fermions.
To distinguish this stage from the subsequent
stages when  particles  decay or interact,
we  call it {\it pre-heating}.
The second stage  is the evolution of the system  after preheating.
During this stage the particles produced from the preheating, can
decay into other particles and self-interact.
 The last stage is the stage of thermalization,
 which can be described by
standard methods.

\section{Back Reaction of Created  Particles}

Creation of particles leads to the several effects
which can change  the dynamics of the system. In particular, it
terminates the broad  resonance particle production.
First,
 the energy from the homogeneous field $\phi (t)$ is transferred
 to the created
particles. The amplitude of the classical oscillations $\phi_0$
is therefore decreasing faster than it would decrease due to the
expansion of the universe only. Second, as it follows from (\ref{35}),
fluctuations contribute to the effective frequency of
the inflaton oscillations:
  $\omega^2_{eff}=m^2_{\phi}+g^2\langle\chi^2
\rangle$ or $ \omega^2_{eff}= c^2\lambda \phi^2 +
3\lambda \langle\ \delta \phi^2 \rangle$.
Third, fluctuations also can contribute to the effective mass of
fluctuations themselves.
As a result, the $q$ parameters in the Mathieu equation are changing,
 in the different models correspondingly as
$q = g^2\phi_0^2/4\omega_{eff}^2$ and
$q= 3\lambda a^2 \phi_0^2/4\omega_{eff}^2$.
The $A(k)$ parameters are changing as
$A(k)=k^2/\omega_{eff}^2a^2 +2q$,
and
 $A(k)=(k^2 + 3\lambda a^2 \langle\ \delta \phi^2 \rangle)
/\omega_{eff}^2 +2q$.
 It  moves the resonance momenta $k$ out the their
initial positions on the instability chart.
The parameter $q$ is decreasing, which  dynamically shifts
 the parameters  towards the narrower
resonance.

As a result, the   creation of particles in the broad resonance regime
is terminated. A simple criterion  of the end of the preheating
is the condition that the energy density of produced fluctuations
 is of the same
order as the  energy density
 the scalar field at the end of preheating $t_{ph}$. For the first model
  $\langle\chi^2\rangle \sim  \phi_0^2$, for the second model
 $\langle\ \delta \phi^2 \rangle  \sim \phi_0^2$.
 From this the  timing of the preheating $t_{ph}$ given above was estimated.

After  copious particle creation around the resonance mode $E$
slowed down, the inflaton field still continues to transfer its energy into
the energy of fluctuations.
Son (1996) noticed that self-interaction of the
$\delta \phi$ particles in the $\lambda \phi^4$ theory might be
important since they have very large occupation number density
$n_E \sim 1/\lambda$ at the resonance mode.
Indeed, the usual estimation for the
 relaxation time $\tau$
 for  the diluted gas of the
 $\delta \phi$ particles
 due to the self-interaction  has limited
relevance when
 the  occupation number density of particles is large.
The scattering of created particles, however,
does not eliminate the preheating regime. The particles created from the
 resonance, in the momentum space occupy only very narrow shells
corresponding to successive resonance bands. Most of the particles after
preheating are located in the first shell of the radius $E$ and the width
$\sim 0.1 E$ (additionally their occupation number $n_k$ is sharply
decreases towards the edges of the shell).
 Two particles from this shell can be scattered inside
 the shell. The rate of redistribution of particles within the shell
is indeed  much larger than $\tau^{-1}$, but it does not terminate the
resonance effect since particles remain inside the resonance shell.
Two particle from the shell can be scattered outside of the resonance shell.
However, this process is much slower than the preheating time $t_{pr}$.
There are interesting possibilities that some of the particles from the first
resonance shell are scattered into other resonance shells. However,
the whole process of rescattering and further creation of particles
in the narrow resonance  after
preheating is rather complicated.
Khlebnikov \& Tkachev (1996) performed first numerical simulation of the
nonlinear evolution of the inflaton field in $\lambda \phi^4$ theory
with rescattering. Based on the result of
Polarski \& Starobinsky (1996)
that the particles created with the large occupation number
can be viewed as  the superposition
of the classical waves with random phases, they modelled
preheating and scattering of $\phi$ particles as the evolution of
classical random gaussian field  with the  spectrum which has
a sharp spike with the amplitude (\ref{occup2}) at the mode $E$
superimposed with the homogeneous background.
Numerical simulation confirms the features of preheating
derived analytically. They also show  the further growth
of the fluctuations $\langle \delta \phi^2 \rangle$
and reveal its spectrum  after the preheating in
 this toy model. We shall note that
in the  more general case, when inflaton field can decay into other bosons,
say,  due to $g^2 \phi^2 \chi^2$-interaction,
the  rescattering is less significant.
We also  do not find evidence of the strong Bose condensations at $k=0$.

\section{Non-Thermal Phase Transitions from After Inflation }

As we have already seen, the reheating scenario depends on the type and
strength of interactions of the involved fields.
In this section we show that the reheating also strongly depends on the
structure of  the elementary particle theory.
So far we considered theories with no symmetry breaking.
In  the case with symmetry breaking,
 in the
beginning, when the amplitude of oscillations $\phi_0$ is much greater than
$\sigma$, the theory of  decay of the inflaton field is the same as in the case
considered above. The most important part of  pre-heating occurs at this stage.
When the amplitude of the oscillations becomes smaller than
$m_\phi/\sqrt\lambda$ and the field begins oscillating near the minimum of the
effective potential at $\phi = \sigma$, particle production due to
the narrow parametric
resonance typically becomes  weak.
The main reason for this is related to the backreaction of
particles created at the
preceding stage of pre-heating on the rate of expansion of the universe and on
 the shape of the effective potential. However, importance of
spontaneous symmetry
breaking for the theory of reheating should not be underestimated, since it
gives rise to the interaction term   $g^2\sigma\phi\chi^2$
or  $\lambda \sigma \phi \delta \phi^2$
 which is linear in
$\phi$. Such terms are necessary for a complete decay of the inflaton field in
accordance with the perturbation theory.

However, in the theories where
preheating is possible one may expect  many  unusual phenomena.
One of the most
interesting  effects  is the
possibility of specific non-thermal post-inflationary phase
transitions which occur after preheating
(Kofman, Linde \& Starobinsky 1996a, Tkachev 1996).
These phase transitions in certain cases can be much more
pronounced
that the standard high temperature cosmological phase transitions
( Kirzhnits \& Linde 1972, Linde 1990). They may lead to copious
production of topological defects and even to a
secondary stage of  inflation after  reheating.

 Let us first remember the  theory of  phase transitions in
theories
with spontaneous symmetry breaking.
For simplicity let us consider  the theory of scalar
fields
$\phi$  the effective potential
\begin{equation}\label{p1}
V(\phi,\chi) =   {\lambda\over2}
(\phi^2-\sigma^2)^2  \ .
\end{equation}
 $V(\phi)$ has a minimum at $\phi =
\sigma$,
 and a maximum at $\phi =  0$ with the curvature
$V_{\phi\phi}
= -m^2= - \lambda\sigma^2$. This effective potential
acquires corrections due  to quantum (or
thermal)
fluctuations of the scalar fields
$
\Delta V =  {3\over 2} \lambda  \langle (\delta\phi^2)\rangle \phi^2 \ .
$
 In
the large temperature limit
$
\langle (\delta\phi)^2\rangle  = {T^2\over 12}.
$
The effective mass squared of the field $\phi$
\begin{equation}\label{p4}
m_{\phi ,eff}^2 = -m^2 + 3 \lambda \phi^2 +
3\lambda \langle
(\delta\phi)^2\rangle
\end{equation}
becomes positive and symmetry  is restored  (i.e. $\phi =0$ becomes
the stable equilibrium point) for $T > T_c$,
where
$T^2_c = {4 m^2\over \lambda } \gg m^2$. At this temperature
the
energy density of the gas of ultrarelativistic particles
is given by
$
\rho = g_*(T_c) {\pi^2\over 30} T_c^4 = {8\, m^4N(T_c)\pi^2\over 15\,
\lambda^2} \
{}.
$

The theory of   cosmological phase transitions is an important part
of the
theory of the evolution of the universe, and during the last twenty
years it was investigated in a very detailed way. However, typically
it
was
assumed that the phase transitions occur in the state of thermal
equilibrium.
Now  we are going to show that similar phase transitions
may occur
even much more efficiently  prior to  thermalization, due to the
anomalously
large expectation values $\langle(\delta\phi)^2\rangle$
produced during the first stage of reheating after
inflation.

We will first consider the model (\ref{p1})
 with the amplitude of spontaneous symmetry breaking $\sigma \ll
M_p$.
As we seen in the previous sections, during preheating
inflaton oscillates
 transfer most of its energy $\sim \lambda
M_p^4$ to
its long-wave fluctuations $\langle(\delta \phi)^2\rangle \sim M_p^2$
 in the
regime
of broad parametric resonance.

The crucial observation   is the following. If the initial energy
density $\sim
\lambda M_p^4$ were instantaneously thermalized, the reheating
temperature
$T_r \sim \lambda^{1/4} M_p$  would be much greater than the typical
particle energy after preheating $E_{\phi} \sim
\sqrt\lambda
M_p$,  and the magnitude of fluctuations
$\langle(\delta\phi)^2\rangle
\sim T_r^2/12 \sim \sqrt \lambda
M_p^2$ would be much smaller than the   magnitude of non-thermalized
fluctuations  $\langle(\delta\phi)^2\rangle\sim M_p^2$. Thus after
the first
stage
of reheating, the
non-thermalized
fluctuations of the scalar field $\phi$ are much greater than the
thermalized
ones. Thermal fluctuations would lead to symmetry restoration in our
model only
for $\sigma {\
\lower-1.2pt\vbox{\hbox{\rlap{$<$}\lower5pt\vbox{\hbox{$\sim$}}}}\
}T_r \sim
\lambda^{1/4} M_p$.   Meanwhile, according to eq. (\ref{p4}),   the
non-thermalized
fluctuations $\langle(\delta\phi)^2\rangle\sim M_p^2$  may lead to
symmetry
restoration in our model even if
the symmetry
breaking parameter $\sigma$  is as large as $M_p$. Thus,  the
non-thermal
symmetry restoration occurs even in those theories where the symmetry
restoration due to  high temperature effects would be
impossible.
Later on, $\langle(\delta\phi)^2\rangle \propto a^{-2},~E_{\phi}
\propto a^{-1}$ because of the expansion of the universe (as far as
$E_{\phi}\gg m$). This leads
to  the  phase transition  with symmetry breaking at the moment
$t=t_c\sim \sqrt{\lambda}M_pm^{-2}$ when $m_{\phi, eff}=0,~
\langle (\delta \phi)^2\rangle=\sigma^2/3,~E_{\chi}\sim m$. Note
that the homogeneous component $\phi (t)$ at this moment is significantly
less than $\sqrt{\langle (\delta\phi)^2\rangle}$ due to its further decay
after preheating.

The mechanism of symmetry restoration  described above
is very
general; in particular, it explains a surprising behavior
of   oscillations of the scalar field found  numerically  in the
$O(N)$-symmetric model of Boyanovsky {\it et al.} (1995a). It is important that
during the
interval between preheating
and the
establishment of thermal equilibrium the universe could experience a
series of
phase transitions which we did not anticipate before.
For example,  cosmic strings and textures, which
could be an additional source  for the formation of the large scale
structure
 of the universe,  should
have $\sigma \sim 10^{16}$ GeV.
 To produce them by
thermal phase
transitions in our model one should have the temperature  after
reheating greater than $10^{16}$ GeV, which is extremely hard to
obtain (Kofman \& Linde 1987).
  Meanwhile, as we see now, fluctuations produced at preheating
may be quite
 sufficient to restore the symmetry. Then the topological defects
  are produced in the standard way when the symmetry breaks down again.
 In other words, production of
superheavy topological defects  can be easily compatible with inflation.

On the other hand, the topological defect production can be quite dangerous.
For example, the model
(\ref{p1}) of a
one-component real scalar field $\phi$ has a discrete symmetry $\phi
\to - \phi$.
As a result, after the phase transition induced by fluctuations
$\langle(\delta\phi)^2\rangle$ the universe may become filled with
domain
walls
separating phases $\phi = +\sigma$ and $\phi = -\sigma$. This
is expected to lead to a cosmological disaster.
Investigation shows that in the theory (\ref{p1}) with
 $\sigma \ll 10^{16}$ Gev fluctuations destroy the coherent
distribution of the background oscillations of $\phi$ and divide
the universe in an equal number of domains with $\phi=\pm \sigma$,
which leads to the domain wall problem.
This mean that in consistent inflationary models of the type of
(\ref{p1}) one should have either no symmetry breaking or
 $\sigma \ge 10^{16}$ Gev.

Similar effects of the nonthermal phase transitions
occur in the models
 where the symmetry breaking occurs for
fields other than the inflaton field $\phi$ (Kofman, Linde \& Starobinsky
 1996a).
The simplest model  has an effective potential
(e.g., Kofman \& Linde 1987;
Linde 1994):
\begin{equation}\label{p7}
V(\phi,\chi) =   {\lambda\over 4} \phi^4
+ {\alpha\over 4}\Bigl(\chi^2  - {M^2\over \alpha}\Bigr)^2 + {
1\over2} g^2 \phi^2 \chi^2 \ .
\end{equation}
 We will assume here that   $\lambda \ll \alpha < g^2$, so
that
at large $\phi$ the curvature of the potential in the
$\chi$-direction is much
greater than in the $\phi$-direction. In this case at large $\phi$
the field
$\chi$ rapidly rolls toward  $\chi = 0$.
 An interesting feature of such models is the symmetry
restoration for the field $\chi$ for $\phi > \phi_c = M/g$, and
symmetry
breaking when the inflaton field $\phi$ becomes smaller than
$\phi_c$. As was
emphasized by Kofman \& Linde (1987),
 such phase transitions may lead to formation
of
topological defects without any need for high-temperature effects.

Now we would like to point out some other specific features of such
models. If
the phase transition discussed above happens during inflation
  (i.e. if $\phi_c > M_p$ in our model), then no new
 phase transitions occur in this model
after reheating.
However, for $\phi_c \ll M_p$ the situation is much more complicated.
First of
all, in this case the field $\phi$ oscillates with the initial
amplitude $\sim M_p$ (if $M^4 < \alpha \lambda M_p^4$). This means
that each time when the absolute value of the
field becomes smaller than $\phi_c$, the phase transition with
symmetry
breaking occurs and topological defects are produced. Then the
absolute value
of the oscillating field $\phi$
again becomes greater than $\phi_c$, and symmetry restores again.
However, this regime does not continue for
a too long time. Within several oscillations, quantum fluctuations of the
field
$\chi$ will be generated with the dispersion
$\langle(\delta\chi)^2\rangle \sim g^{-1}\sqrt {\lambda}M_p^2$.
For $M^2<g^{-1}\sqrt{\lambda}\alpha M_p^2$,
these fluctuations will keep the symmetry restored. The symmetry breaking
will be
finally completed  when $\langle(\delta\chi)^2\rangle$ will become
small enough.
Riotto and Tkachev (1996) noted that in the model   (\ref{p7})
in  the case of the strong self-interaction of
 $\chi$ particles,
  $\lambda \ll g^2 \ll \alpha$,  the presence of a
huge occupation number of the created $\chi$ particles $n_E \sim 1/g^2$
would require  the consideration beyond
the one-loop approximation. It can be shown, however, that
in this case the the preheating stops earlier when  $n_E \sim 1/\alpha$
(Kofman, Linde \& Starobinsky 1996b).

One may imagine even more complicated scenario when oscillations
of the
scalar field $\phi$ create large fluctuations of the field $\chi$,
which in
their turn interact with the scalar fields $\Phi$ breaking symmetry
in GUTs.
Then we would have phase transitions in GUTs induced by the
fluctuations of the
field $\chi$. Note that in the models considered in this section the
field
$\chi$ does not oscillate near $\chi = 0$ prior to the phase
transition, since
such oscillations are damped out during the long stage of inflation
prior to
the phase transition.  Thus oscillations of the field $\chi$ in the theory
(\ref{p7}) definitely do not suppress  the topological defect production.
This means that no longer can the absence of primordial monopoles be
considered as
an automatic consequence of inflation. To avoid the monopole production one
should  use the theories where   quantum fluctuations produced during
preheating are small or decoupled from the GUT sector. This condition imposes
additional constraints on realistic inflationary models.

\section{Conclusions }

 We have found that the decay of the inflaton field
typically begins with the  explosive production of particles during the
 stage
of preheating in the regime of  a broad parametric resonance.
 During the second stage,
 the inflaton field decays further in the regime of the narrow resonance,
and  particles produced at this
stage  decay into other particles and self-interacting.
The third stage is the thermalization.
The two last stages require  much more time that the stage of preheating.
Typically, coupling constants of interaction of the
inflaton field with matter are extremely small, whereas coupling constants
involved in the decay of  other particles  can be much greater.
As a result,  the reheating temperature $T_r$ given by (\ref{12})
will   not necessarily be defined by the perturbative decay rate
of the inflaton oscillations (\ref{7}).
$T_r$ can be much higher than the typical temperature $T_r
{\ \lower-1.2pt\vbox{\hbox{\rlap{$<$}\lower5pt\vbox{\hbox{$\sim$}}}}\ } 10^9$
GeV,
which could be obtained
neglecting the stage of parametric resonance.

A specific feature of the preheating is that bosons produced
at this stage are far away from thermal equilibrium and typically
have enormously large occupation numbers. This may have very interesting
applications to cosmology.
There is a   new class of phase transitions that
 may occur
 at the intermediate
stage between the end of inflation and the establishment of thermal
equilibrium. These phase transitions
 take place  even in the theories
where the
scale of spontaneous symmetry breaking is comparable to $M_p$ and where the
reheating temperature is very small.
Therefore,
phase transitions of the new
type may
have dramatic consequences for inflationary models and the theory
 of physical processes in
the very
early universe.

The preheating may help with the baryogenesis problem
(Kofman, Linde \& Starobinsky 1996a; Yoshimura 1996).
  In the models of GUT baryogenesis, it was assumed
that  GUT symmetry was restored by high-temperature effects,
since otherwise the density of X, Y, and superheavy Higgs bosons would
 be very small. This condition is hardly compatible with inflation.
It was also required that the products of decay of these particles
should stay out of thermal equilibrium, which is
a very restrictive condition. In our case, the superheavy particles
 responsible for baryogenesis can be abundantly produced by parametric
resonance, and they as well as the  products of their decay will be
out of thermal equilibrium until
 the end of reheating.

 Another  consequence of the resonance effects is a fast
 change of the equation of state from a vacuum-like one to the
equation of state of relativistic matter. This leads to
suppression of    primordial black holes formation that could be produced
after inflation.

The preheating stage may appear not only in the models where the reheating
 occurs due to the
decay of the homogeneous scalar field. Kolb \& Riotto (1996) suggested that
preheating may also occur  in first order inflation.
It would be very interesting to investigate this problem, as well as the
 many other exciting possibilities  which  the new theory
of reheating may offer us.

\end{document}